\documentclass[12pt]{article}
\usepackage{graphicx}
\begin{document}
\begin{center} {\large \bf Spin-Dependent Transport through the Finite Array of
Quantum Dots: Spin Gun}
\end{center}

S.A.Avdonin$^{1)}$, L.A.Dmitrieva$^{2)}$, Yu.A.Kuperin$^{3)}$, V.V.Sartan$^{3)}$

\vskip0.2cm
{\small \begin{center}
1) Department of Mathematical Sciences,\\
University of Alaska Fairbanks,
Alaska 99775-6660, USA\\
ffsaa@uaf.edu
\vskip0.2cm
2) Division of Mathematical and Computational Physics,\\
St.Petersburg State University,
198504 St.Petersburg, Russia\\
mila@JK1454.spb.edu

\vskip0.2cm
3) Laboratory of Complex Systems Theory,\\
St.Petersburg State University,
St.Petersburg 198504, Russia\\
kuperin@JK1454.spb.edu, vsartan@hotmail.com
\end{center}}

\vskip0.3cm
\begin{abstract}
The problem of spin-dependent transport of electrons through a
finite array of quantum dots attached to 1D quantum wire (spin
gun) for various semiconductor materials is studied. The
Breit-Fermi term for spin-spin interaction in the effective
Hamiltonian of the device is shown to result in a dependence of
transmission coefficient on the spin orientation. The difference
of transmission probabilities for singlet and triplet channels can
reach few percent for a single quantum dot. For several quantum
dots in the array due to interference effects it can reach
approximately $100\%$ for some energy intervals. For the same
energy intervals the conductance of the device reaches the value
$\approx 1$ in $[e^{2}/\pi\hbar]$ units. As a result a model of
the spin-gun which transforms the spin-unpolarized electron beam
into completely polarized one is suggested.
\end{abstract}
\section{Introduction}
Spintronics is a novel branch of nanoscience which exploits spin
properties of electrons or nuclei instead of charge degrees of
freedom \cite{Asholm}. Perhaps the most current efforts in
designing and manufacturing the spintronics devices are focused
now on finding novel ways of both generation and utilization of
spin-polarized currents. This, in particular, includes the study
of spin-dependent transport in semiconductor nanostructures, which
can operate as spin polarizers or spin filters. Although
spintronics can have a lot of practical applications \cite{coll0}
- \cite{coll1}, one of them is most interesting and ambitious: the
application of electron or nuclear spins to quantum information
processing and quantum computation \cite{comp0} - \cite{comp1}.

In this paper we suggest a new type of spin polarizer constructed from the finite
array of semiconducting quantum dots (QD) attached to a quantum wire (QW).

\begin{figure}[h]
\includegraphics[width=14cm] {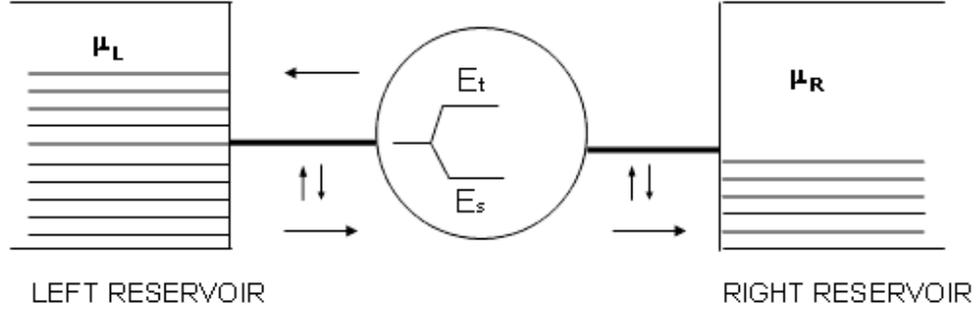}
\caption{ \small {Schematic representation of one quantum dot
attached to quantum wires. The left and right reservoirs are at
different chemical potentials \protect $\mu_L$ and $\mu_R$. The
horizontal arrows represent incident, transmitted and reflected
electronic wave. The quantum dot carries spin $s=\frac{1}{2}$. By
$E_{s}$ and $E_{t}$ the singlet and triplet states are denoted. By
$\uparrow\downarrow$ the electron spin projection
$s_{3}=\pm\frac{1}{2}$ before and after the scattering by the
quantum dot is denoted. If the small voltage $eV =\mu_L- \mu_R$ is
applied a non-equilibrium situation is induced and spin-dependent
current will flow through the device.}}
\end{figure}

\normalsize

\begin{figure}[h!]
\includegraphics[width=14cm] {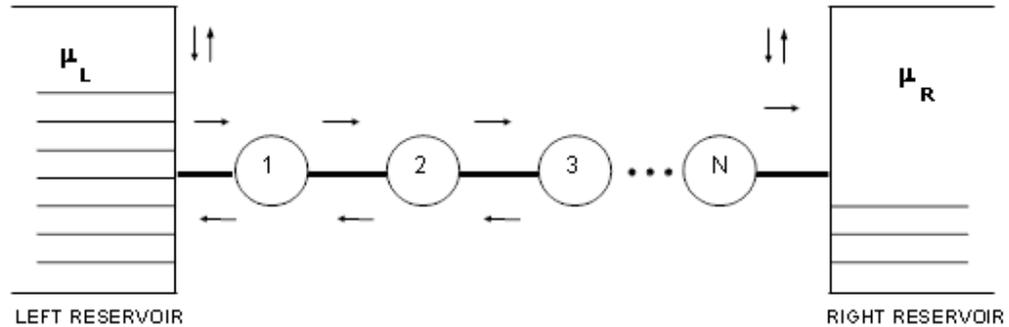}
\caption{ \small {Schematic representation of spin filter device (spin gun)
fabricated from $N$ quantum dots and quantum wires. For details see the caption
for Fig.1  }}
\end{figure}
\normalsize

All dots of the array are identical and each of them carries the spin 1/2. The schematic
construction of the proposed device is shown for a single dot in Fig.1 and for the finite
array of dots in Fig.2.

Using the mathematical modeling based on zero-range potentials
\cite{zero}, \cite{Albev} we reduce the problem of spin-dependent
transport in the device to exactly solvable scattering problem. In
terms of the scattering data, in particular, in terms of
spin-dependent transmission coefficients through the device, we
calculate the most important measurable characteristics: the
polarization efficiency $P(k)$ and the conductance $G(k)$ of the
device.

\section{Assumptions, constrains and choice of model parameters}
In this section we formulate the physical conditions under which
our mathematical modeling is relevant. We assume that in the
quantum wire connected with a quantum dots one-mode propagation of
electrons in a ballistic regime can be realized. For the ballistic
propagation of electrons through the QW the de Broglie wavelength
\cite{Ando} $\lambda_{B}=\frac{2\pi\hbar}{\sqrt{2m^{\ast}k_{B}T}}$
has to be much greater than both  the mean free path $l_{mfp}$ in
the material of QW and the length $d$ of QW between the
neighboring quantum dots, i.e. $\lambda_{B}\gg l_{mfp}$ and
$\lambda_{B}\gg d$.

On the other hand in order to use zero-range potentials for
mathematical modeling of scattering by quantum dots the de Broglie
wavelength has to be greater than the mean size $r_{0}$ of the
dot: $\lambda_{B}\gg r_{0}$ \cite{zero}. There are different ways
to manufacture quantum dots (see, e.g. \cite{Jeff}, \cite{Jacak},
\cite{dot}). Depending on technology the mean size of quantum dots
varies from 20 nm for small dots up to more than 100 nm for the
large dots. In order to satisfy all the constrains mentioned above
we have to assume that spin-gun can be realized on relatively
small QD using narrow-gap semiconductors (i.e. semiconductors with
small $E_g$ --- see Table 1). For such materials (see Table 1)
$\lambda_{B}$ is about 50 nm at room temperature and more than 100
nm at the liquid nitrogen temperature. \vskip0.3cm
\centerline{Table 1.\cite{Anton}}

\begin{center}
\small
\begin{tabular}{|l|c|c|c|c|}\hline
Semiconductor & {\tabcolsep=0em \begin{tabular}{c}\strut $\!\!\!\!E_g,\, eV$\\ (285\,K)
\end{tabular}} & $\frac{m^{\ast}}{m_e}$ &
{\tabcolsep=0em
\begin{tabular}{c} \strut $\lambda_{B} ,\,$nm \\ (285\,K)\end{tabular}} & {\tabcolsep=0em
\begin{tabular}{c} $\lambda_{B} , \,$nm\\ (77\,K)\end{tabular}}
\\ \hline
$GaAs$ &                 1.430  & 0.068  &  29.9 &  57.7 \\ \hline
$InAs$ &                 0.360  & 0.022  &  52.8 & 101.5\\ \hline
$Cd_xHg_{1-x}Te$ & \multicolumn{4}{|l|}{} \\ \hline
$x=0.20$ & 0.150  & 0.013 &  68.6 & 131.9\\ \hline
$x=0.30$ & 0.290  & 0.021 &  54.0 & 103.9\\ \hline
$HgTe$   & -0.117 & 0.012  & 71.4 & 137.4\\ \hline
$Zn_{0.15}Hg_{0.85}Te$ & 0.190  & 0.015  &  63.9 & 122.9\\ \hline
$InSb$ & &0.014 &66.1 & 127.2\\ \hline
\end{tabular}
\end{center}
\normalsize
It means that we have either to use in our modeling the parameters $d, r_{0}\ll$ 100 nm
or to assume the working temperature of the device less than $T=77 K$. Below in Section 5
we shall take into account these limitations in our numerical simulations of
the device operation.

\section {A mathematical model of spin transport through a single quantum dot}

In this section we consider a model for scattering of a 1D
electron on a quantum dot  attached to the quantum wire. Both 1D
electron and quantum dot are assigned to have spin $1/2$
($s_{e}=1/2,\,\, s_{d}=1/2$ and the possible spin projection on a
fixed axis are $(s_{3})_{e,d}=\pm1/2.$ We emphasize that in the
model in question the quantum dot is treated as unstructured
uncharged scatterer carrying spin $1/2$ as a whole.

Due to the Breit-Fermi theory \cite{Breit} the spin-spin
interaction in frames of this simplest model leads to a singular
interaction $V^{s}$ in the model Hamiltonian: $
V^{s}=\hat{\gamma}_{s}\delta(x) \otimes \langle
\hat{s}(1),\hat{s}(2)\rangle, $ where $x$ is the distance between
the electron and the quantum dot, $\delta(x)$ is the 1D delta
function, $\hat{\gamma}_{s}$ is the coupling constant,
$\hat{s}(1)$ and $\hat{s}(2)$ are the electron and the quantum dot
spin operators respectively: $\hat{s}(j) = (s_{1}(j), s_{2}(j),
s_{3}(j)),$ $ j=1, 2,$ $s_{i}=\frac{\hbar}{2}\sigma_{i}$ where
$\sigma_{i}, i=1, 2, 3$ are the Pauli matrices and $\langle
\hat{s}(1), \hat{s}(2)\rangle = s_{1}(1) \otimes s_{1}(2)+s_{2}(1)
\otimes s_{2}(2)+ s_{3}(1) \otimes s_{3}(2).$

Thus, the  Schr\"{o}dinger equation relevant to the proposed model
has the form
\begin{equation}\label{equat}
\left[-\frac{d^{2}}{dx^{2}}\otimes I_{4} +
\gamma_{s}\delta(x)\otimes\langle s(1), s(2)\rangle\right]\,\, \Psi(x,s_{3}(1),s_{3}(2))=
k^{2}\Psi(x,s_{3}(1),s_{3}(2)),
\end{equation}
Here $I_{4}$ is the unit operator in the total spin space
${\mathbf{C}}^{2}\otimes {\mathbf{C}}^{2}$ and $m^{\ast}$ is the
effective mass of electron in semiconductor material of the
quantum wire and $\otimes$ means the tensor product, $k^{2}= E
\frac{2m^{\ast}}{\hbar^{2}}$, $s(j)= \hbar^{-1} \hat{s}(j)$,
$\gamma_{s} = 2 m^{\ast}\hat{\gamma}_{s}$. The  wave function
$\Psi$ depends on the coordinate variable $x$ and
$s_{3}(1),s_{3}(2)$ can take the values $\pm 1/2$.

In accordance with the extensions theory approach \cite{Albev} the equation
(\ref{equat}) is equivalent to the following boundary problem
\begin{equation}\label{Psi}
\left(-\frac{d^{2}}{dx^{2}}\otimes I_{4}\right) \Psi = k^2 \Psi
\end{equation}
$$
\Psi|_{x=+0} = \Psi|_{x=-0}
$$
\begin{equation}\label{bval}
\Psi^{\prime}|_{x=+0} - \Psi^{\prime}|_{x=-0} =
\gamma_{s} \langle s(1), s(2)\rangle \,\Psi|_{x=0}.
\end{equation}

In order to solve the problem (\ref{Psi}), (\ref{bval}) one has to
separate the spin variables, i.e. to calculate the matrix elements
of the Hamiltonian from equation (\ref{equat}) in the spin basis
constructed below. Let us denote by $\chi_{\pm}(1)$ and
$\chi_{\pm}(2)$ the spin part of the electron and  the dot wave
functions respectively with the value of spin projection
$s_{3}(j)=\pm1/2, j=1, 2.$ One can take these spinors in the form
$\chi_{+}(j)= \left(\begin{array}{c}
1\\
0
\end{array}\right),$ $\chi_{-}(j)=
\left(\begin{array}{c}
0\\
1
\end{array}\right)\,.$
It is well known \cite{Faddeev} that the two-spin space
${\mathcal{H}}_{s_{1}}\otimes{\mathcal{H}}_{s_{2}} \simeq
{\mathbf{C}}^{2} \otimes {\mathbf{C}}^{2}$ can be decomposed into
the direct sum of two invariant (with respect to operators of
representation of the group of rotation) subspaces:
${\mathcal{H}}_{s_{1}}\otimes{\mathcal{H}}_{s_{2}} = D_{0} \oplus
D_{1},$ where $D_{0}$ is the singlet space (subspace of the total
spin $S=0$)  which is spanned on the spinor $\chi_{0,0},\,\, \dim
D_{0} = 1$, and $D_{1}$ is the triplet space (subspace of the
total spin $S=1$)  which is spanned on three spinors
$\{\chi_{1,1}, \chi_{1,0}, \chi_{1,-1}\}$, $\dim D_{1} = 3$.

The orthonormal basis of spinors in the triplet subspace
has the form \cite{Faddeev}: $\chi_{1,1}=\chi_{+}(1)\otimes\chi_{+}(2)$, $\chi_{1,-1}=\chi_{-}(1)
\otimes\chi_{-}(2)$, $\chi_{1,0}=\frac{1}{\sqrt{2}}(\chi_{+}(1)\otimes\chi_{-}(2) +
\chi_{-}(1)\otimes\chi_{+}(2))$.
The second index in $\chi_{1,S_{3}}$ means the values of the total spin projection
$S_{3}=1,0,-1$
The basis vector in the singlet subspace reads
$\chi_{0,0}=\frac{1}{\sqrt{2}}(\chi_{+}(1)\otimes\chi_{-}(2) -
\chi_{-}(1)\otimes\chi_{+}(2))$.
The above described spinors form the orthonormal basis in the total spin space
${\mathcal{H}}_{s_{1}}\otimes{\mathcal{H}}_{s_{2}} \simeq
{\mathbf{C}}^{2} \otimes {\mathbf{C}}^{2}$. Hence the total wave function
$\Psi(x,s_{3}(1),s_{3}(2))$ can be decomposed as follows
\begin{equation}\label{wavefunc}
 \Psi(x,s_{3}(1),s_{3}(2))= \psi_{1,1}(x)  \chi_{1,1}+ \psi_{1,0}(x)  \chi_{1,0}+\psi_{1,-1}(x)  \chi_{1,-1}
+\psi_{0,0}(x)  \chi_{0,0}
\end{equation}

In order to separate the spin variables
in the problem (\ref{Psi}), (\ref{bval}) it is sufficient to calculate the action of the
operator $\langle s(1), s(2)\rangle$ on these spinors. One can easily verify that
$\langle s(1),s(2)\rangle\,\chi_{1,S_{3}} =\frac{1}{4}\chi_{1,S_{3}}$, where
$S_{3}=1,0,-1$ and
$\langle s(1),s(2)\rangle\,\chi_{0,0} =-\frac{3}{4}\chi_{0,0}$.

Since the chosen basis of spinors is an orthonormal system
the separation of spin variables in the channel $S=0$ and $S=1 $
gives the following boundary value problem for the coordinate parts
of the wave function (\ref{wavefunc})
\begin{equation}\label{psiunite}
(-\frac{d^{2}}{dx^{2}}\otimes I_{2}-k^{2})\,\,\psi\,\,=0
\end{equation}
$$
\psi(+0) = \psi(-0)
$$
\begin{equation}\label{bvalunite}
\psi^{\prime}|_{x=+0} -\psi^{\prime}|_{x=-0} =
\left(\begin{array}{cc}
\gamma_{11}&0\\
0&\gamma_{00}
\end{array}\right)\,\, \psi(0).
\end{equation}
Here $\psi$  is a two-component vector ${\bf{\psi}}=
\left(\begin{array}{c}
\psi_{1, S_{3}}\\
\psi_{0,0}
\end{array}\right)(x)$ where $S_{3}=1,0,-1$, $\gamma_{00}= -\frac{3}{4}\gamma_{s}$
is the coupling constant in the singlet channel and $\gamma_{11}= \frac{1}{4}\gamma_{s}$
is the coupling constant in the triplet channel. In the proposed model in the absence
of relativistic effects the total spin $S$ conserves and the triplet-singlet and
singlet-triplet transitions are forbidden. As a consequence, the antidiagonal
elements of the coupling matrix in boundary condition (\ref{bvalunite}) are equal
to zero. Since the projection of the total spin $S_{3}$ also conserves, one can choose
in the boundary problem above any value of this projection. In what follows we set
$S_{3}=1$.

\section{Scattering on periodic array of $N$ quantum dots}
The single-dot Hamiltonian (\ref{Hamilt})
for the case of periodic array of $N$ quantum dots is modified as follows
\begin{equation}\label{Hamilt}
H^{(N)} = -\frac{\hbar^{2}}{2m^{\ast}}\frac{d^{2}}{dx^{2}}\otimes
I_{4} +
\sum_{l=1}^{l=N}\hat{\gamma}_{s}\delta(x-y_{l})\otimes\langle
\hat{s}(1),\hat{s}(2)\rangle,
\end{equation}
where points $y_{l}$ are separated by  distance $d$.

All the constructions of the previous section are easily
generalized for this case. Hence after separation of spin
variables in the equation $H^{(N)}\Psi^{(N)} = E\Psi^{(N)}$ one
obtains for the coordinate part
 $\psi^{(N)}(x)=
\left(\begin{array}{c}
\psi_{1, 1}^{(N)}\\
\psi_{0,0}^{(N)}
\end{array}\right)(x)$  of the total wave  function $\Psi^{(N)}$ the
following boundary
value problem
\begin{equation}\label{psiuniteN}
(-\frac{d^{2}}{dx^{2}}\otimes I_{2}-k^{2})\,\,\psi^{(N)}\,\,=0
\end{equation}
$$
\psi^{(N)}(y_{l}+0) = \psi^{(N)}(y_{l}-0)
$$
\begin{equation}\label{bvaluniteN}
{\psi^{(N)}}^{\prime}|_{x=y_{l}+0} -{\psi^{(N)}}^{\prime}|_{x=y_{l}-0} =
\left(\begin{array}{cc}
\gamma_{11}&0\\
0&\gamma_{00}
\end{array}\right)\,\, \psi^{(N)}(y_{l}),\,\, l=1,2,...,N.
\end{equation}

The transmission coefficient in the triplet channel
$T_{11}^{(N)}(k)$  and singlet channel $T_{00}^{(N)}(k)$  are
calculated in terms of elements of the ${4\times4}$ matrix
$\mathbf{ \hat T} = {t_{ij}}$, $i,j=1,2,3,4$, ($i$ is the line
number, $j$ is the column number) which is the product of
transition matrices through dots and wires connected dots. It
reads $\mathbf{\hat T} = Q^{N-1}Q_{1}$, where
\begin{equation}\label{matrix}
Q=\frac{1}{2ik}
\left(\begin{array}{cccc}
\gamma_{11}+2ik&0&e^{-2idk}\gamma_{11}&0\\
0 & \gamma_{00}+2ik&0&e^{-2idk}\gamma_{00}\\
-\gamma_{11} & 0&e^{-2idk}(2ik-\gamma_{11})&0\\
0 & -\gamma_{00}&0&e^{-2idk}(2ik-\gamma_{00})
\end{array}\right).
\end{equation}
The matrix $Q_{1}$ differs from $Q$ by absence of multiplier
$e^{-2idk}$ in its matrix elements. Here $d$ is the distance
between the centers $y_{l}$ where $\delta-$functions are
localized. As applied to the physical situation we are modeling
$d$ is the length of a quantum wire connecting two neighbor dots.
The transmission coefficients have the form
$T_{11}^{(N)}=t_{11}+t_{13}e^{-2iky_{1}}R_{11}^{(N)}$,
$T_{00}^{(N)}=t_{22}+t_{24}e^{-2iky_{1}}R_{00}^{(N)}$. The
reflection coefficients $R_{11}^{(N)}$, $R_{00}^{(N)}$ in the
triplet and singlet channels respectively are given by equations
$R_{11}^{(N)} =e^{2iky_{1}}D^{-1}(t_{41}t_{34}-t_{31}t_{44})$,
$R_{00}^{(N)} =e^{2iky_{1}}D^{-1}(t_{43}t_{32}-t_{33}t_{42})$,
where $D=t_{33}t_{44}-t_{34}t_{43}$. The relations
$|T_{jj}^{(N)}|^{2}+|R_{jj}^{(N)}|^{2}=1$ \  ($j=0,1$) which imply
the unitarity of scattering has been checked by means of
analytical calculations package.

In the case $N=1$ one has $T_{jj}^{(1)} =
2ik(2ik-\gamma_{jj})^{-1}$, $R_{jj}^{(1)} =
e^{2iky_{1}}\gamma_{jj}(2ik-\gamma_{jj})^{-1}$ ($j=0,1$)

\section{Results and discussions}

In this section we present the results of the numerical
calculations of the operation regime characteristics for the spin
gun and discuss the proper choice of the model parameters as well
as the choice of semiconductor materials for QW and QD. We use the
applied bias $V:\,\,eV = \mu_{L} - \mu_{R}$ for tuning the spin
polarization effects in the device.

In order to calculate the conductance $G$ of the device we exploit
the two-channel Landauer formula at zero-temperature limit:
$$
G = \frac{e^{2}}{\pi\hbar}\left[|T_{00}^{(N)}(E_{F}+eV)|^{2} +
|T_{11}^{(N)}(E_{F}+eV)|^{2}\right],
$$
where $E_F$ is the Fermi level of QW.

For unpolarized incident beam of electrons the scattering by QD array in the
triplet and singlet channels leads to different transmission coefficients
$T_{11}^{(N)}$ and $T_{00}^{(N)}$. Following \cite{Perel} we shall use the
polarization efficiency
$$
P =\frac{|T_{11}^{(N)}|^{2}-|T_{00}^{(N)}|^{2}}{|T_{11}^{(N)}|^{2}+
|T_{00}^{(N)}|^{2}}
$$
which determines the difference of  transmission
probabilities through the spin gun for the triplet and  singlet spin states.

At a given wave number $k$ the conductance $G$ and polarization
efficiency $P$ of the device both depend on the model parameters:
the distance $d$, the coupling constants $\gamma_{11}$,
$\gamma_{00}$  and the number $N$ of QD. They also both depend on
the choice of semiconductor materials from which the QD and QW are
fabricated, i.e. effective mass $m^{\ast}$, Fermi energy $E_{F}$
and mean size $r_{0}$ of quantum dots. Varying the bias applied
one can calculate the dependence of $G$ and $P$ on $k=
\sqrt{\frac{2m^{\ast}(E_{F}+eV)}{\hbar^{2}}}$ at fixed model
parameters and chosen $m^{\ast}$, $E_{F}$ and $r_{0}$. Let us
mention that from spin analysis in Section 3 one has $\gamma_{00}
= -3 \gamma_{11} $. Since $d$ and $N$ are free parameters of the
model the only parameter has to be fixed is $\gamma_{11}$. The
reasonable way to fix this parameter is to take it proportional to
the inverse mean size of the quantum dot. For relatively small QD
$r_{0}\approx 20 \div30\, nm$. Hence $\gamma_{11} \approx 0.03
\div 0.05 nm^{-1}$. Taking $d \approx 50 nm,$ we obtain the value
of order $\approx 1$
 for the dimensionless parameter
${\hat\gamma}_{11} = \gamma_{11}\frac{d}{\pi}.$

In Figures 3 - 5 we show the dependence of $G$ and $P$ on the
dimensionless parameter $\hat{k} = k \frac{d}{\pi}$ for
$\gamma_{11} \frac{d}{\pi} = 1$  in the cases $N = 1, 2, 8$. We
also indicate the position of Fermi level $E_{F} = 35\, meV$ in
$InSb$ ($m^{\ast}= 0.014 m_{e}$) and fix the inter-dot distance $d
= 45\, nm$.

The general behavior of the conductance $G$ and polarization
efficiency $P$ for the wide range of parameters variation is the
following. On the $\hat{k}$-axis with the growth of the number $N$
of quantum dots there arise the so-called ``working windows'' (see
Figure 5), i.e. such intervals $[{\hat{k}}_{1},\, {\hat{k}}_{2}]$,
$[{\hat{k}}_{3},\, {\hat{k}}_{4}]$,...in which $P$ and $G$ are
sufficiently high simultaneously. Due to the interference
processes in a QD array at $N \sim 8 \div 10$ the values of $P$
and $G$ on the working windows approach $P\approx 100\%$ and $G
\approx \frac{e^{2}}{\pi\hbar}$ which equals to one half of its
upper limit. It means that at $\hat{k}$ lying in the working
windows the spin gun really produces the fully polarized spin beam
and conducts at fairly high level. In Figures 3 - 5 the working
windows lie on the right of $E_F$ up to the value of ${\hat k} =
\frac{d}{\pi}\sqrt{\frac{2m^{\ast}(E_{F}+eV)}{\hbar^{2}}}$ which
corresponds to the bias applied.

\section{Plans and prospect}
There are a lot of interesting modifications of the model we
constructed. For example, we are going to take into account the
internal structure of quantum dots. We plan to supply the quantum
dot by prescribed internal structure remaining the model exactly
solvable. Other interesting modification is to take into account
the Coulomb interaction between the electron and quantum dot if
the dot is charged. We shall prove that if we switch on the
Coulomb interaction the modified model remains exactly solvable.
Finally we are going to incorporate in our model the magnetic
field in order manipulate spins in QW and QD.

\section{Acknowledgments}
The idea to use interference effect in finite periodic structures in order to
amplify the polarization of electronic beam and the name of a such type
spintronic devices (spin-gun) were proposed by N.T. Bagraev and
B.S. Pavlov {\footnote {unpublished manuscript}}. The authors are very grateful
to B.S. Pavlov for supplying the manuscript and for numerous fruitful
discussions.

\begin{figure}[h!]
\includegraphics[width=11cm] {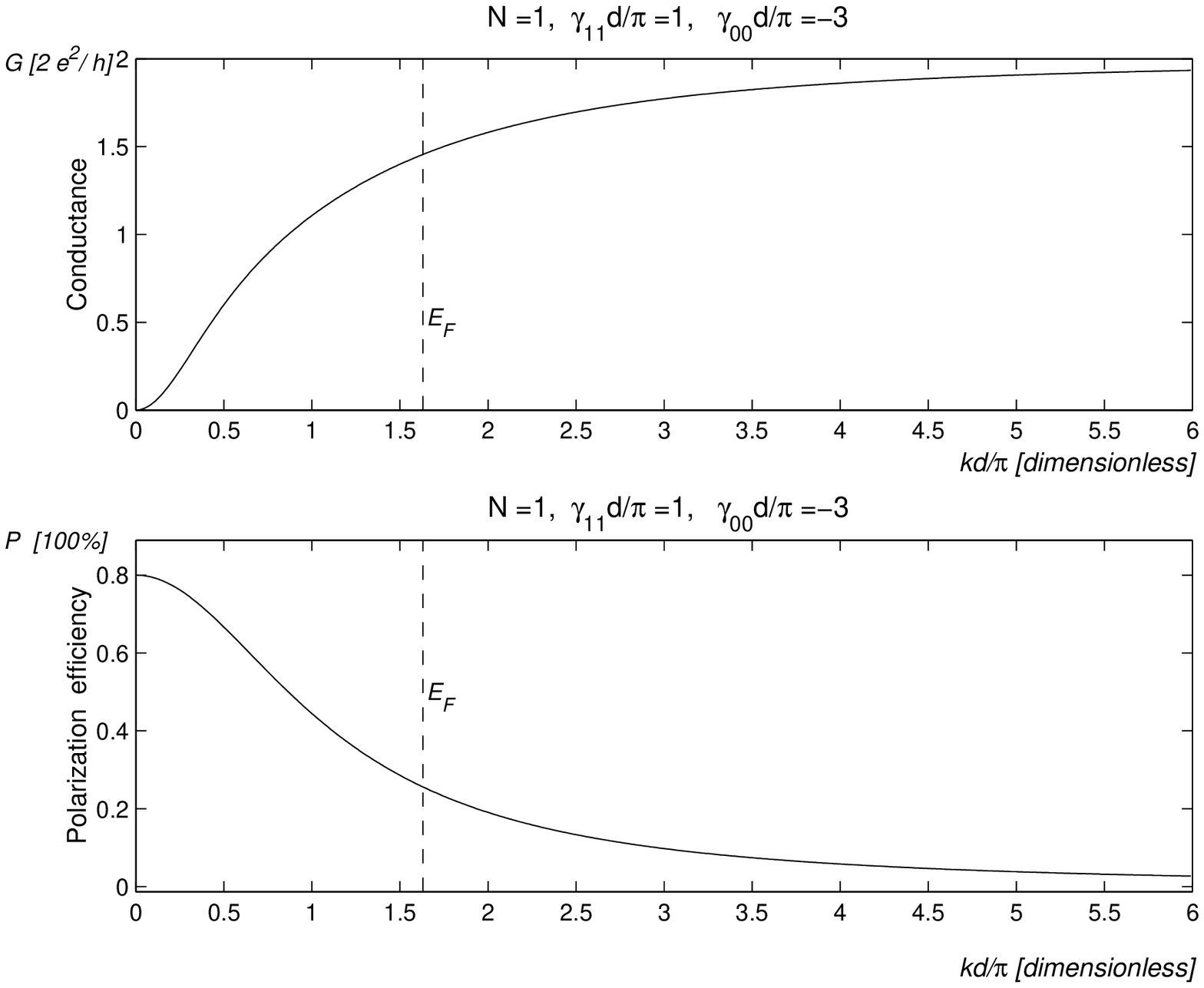}
\caption{ \small {The conductance $G$ and polarization efficiency
$P$ versus dimensionless momentum $\hat{k} = k \frac{d}{\pi}$ of
incident electrons in the case of one quantum dot. By the dashed
line the Fermi level  $E_{F} = 35\, meV$ in $InSb$ is indicated.
The interdot distance $d = 45\, nm$.}}
\end{figure}
\normalsize
\begin{figure}[h!]
\includegraphics[width=11cm] {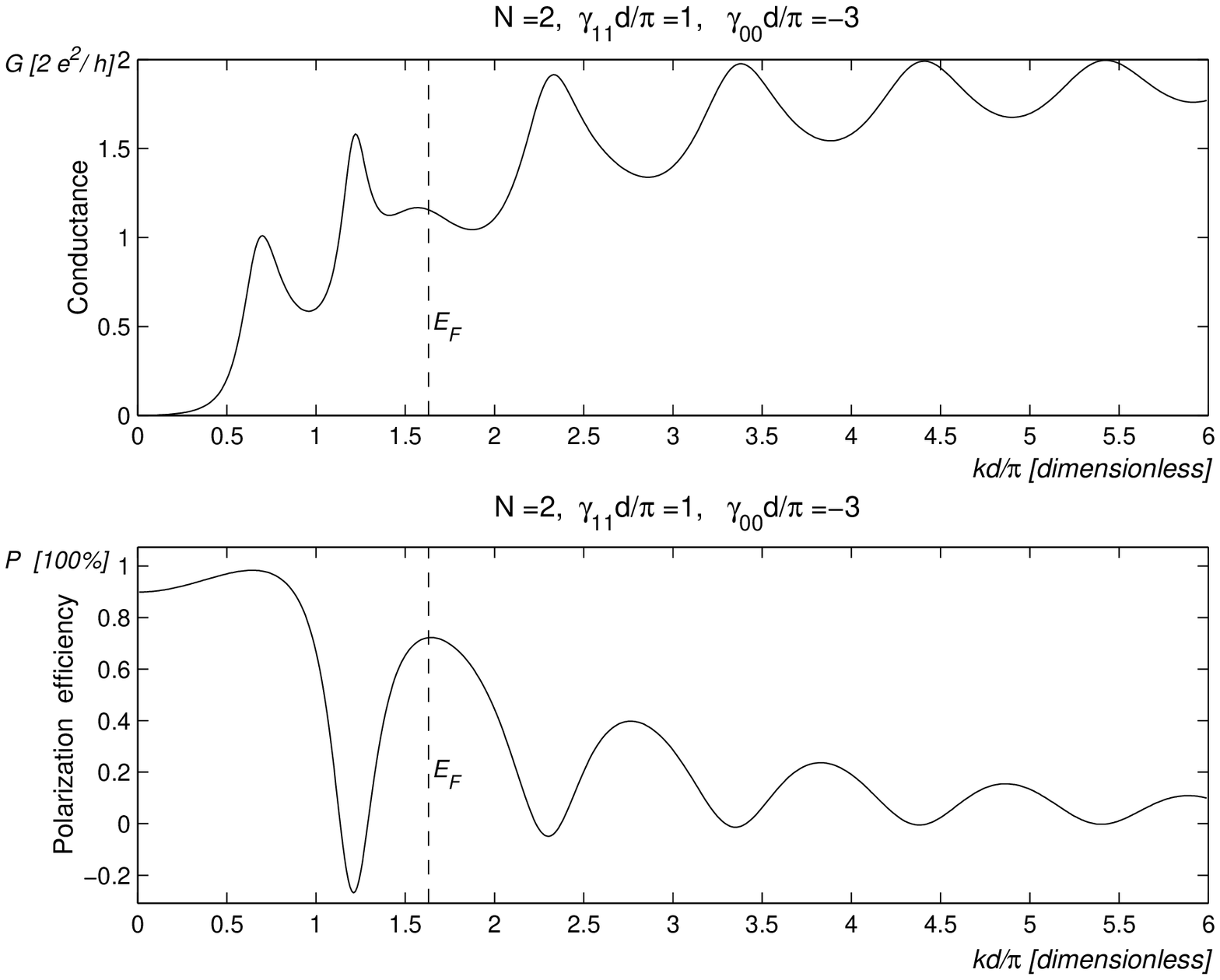}
\caption{ \small {The conductance $G$ and polarization efficiency
$P$  versus dimensionless momentum $\hat{k} = k \frac{d}{\pi}$ of
incident electrons in the case of two quantum dots. By the dashed
line the Fermi level  $E_{F} = 35\, meV$ in $InSb$ is indicated.
The interdot distance $d = 45\, nm$.}}
\end{figure}
\normalsize
\begin{figure}[h!]
\includegraphics[width=11cm] {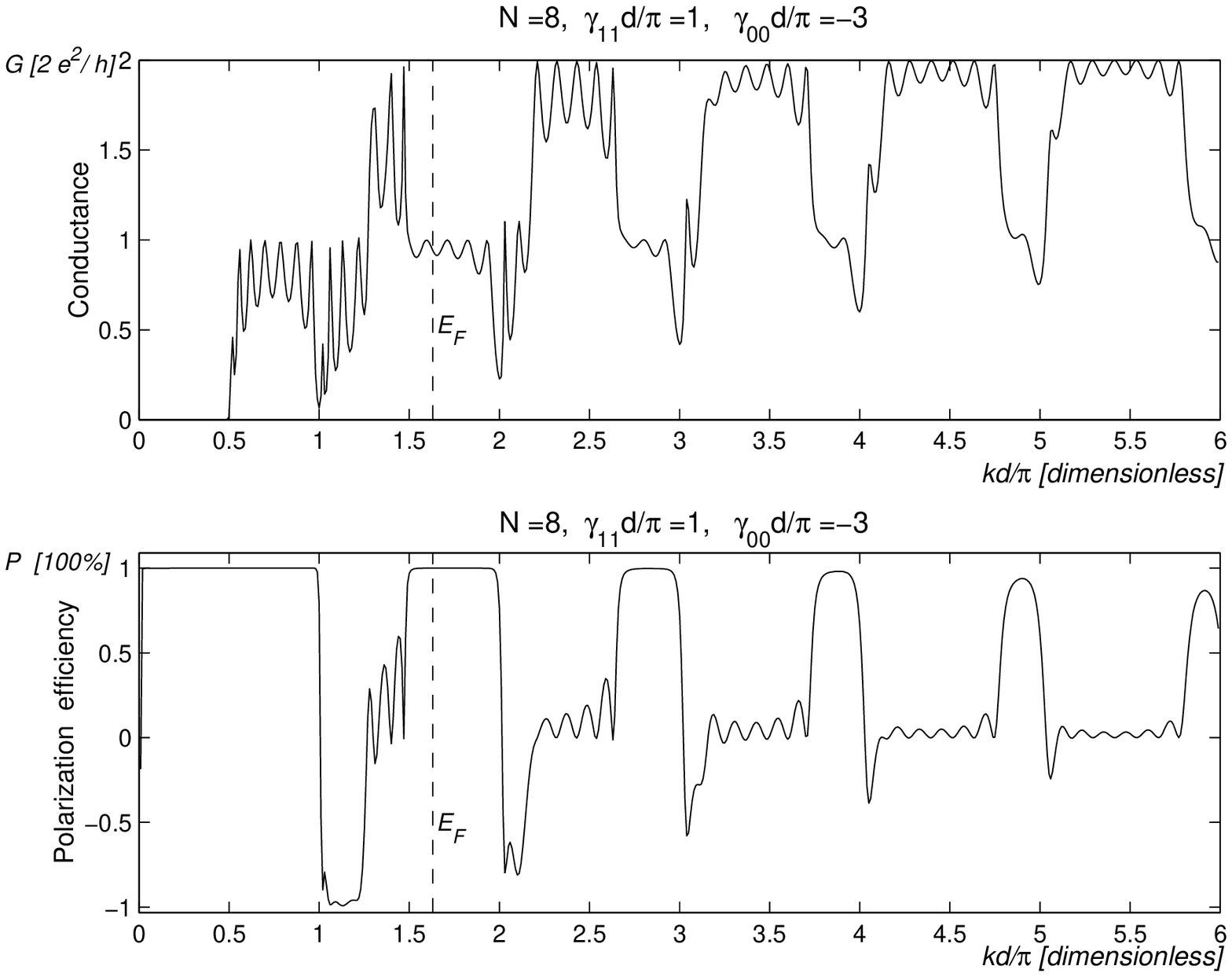}
\caption{ \small {The conductance $G$ and polarization efficiency
$P$ versus dimensionless momentum $\hat{k} = k \frac{d}{\pi}$ of
incident electrons in the case of 8 quantum dots. By the dashed
line the Fermi level  $E_{F} = 35\, meV$ in $InSb$ is indicated.
The interdot distance $d = 45\, nm$.}}
\end{figure}
\normalsize

\end{document}